\documentclass[12pt,a4paper]{article}
\usepackage{graphicx}
\begin{document}
\textwidth=135mm
 \textheight=200mm
\begin{center}
{\bfseries Can $\eta$ mesons have a magnetic moment?
}
\vskip 5mm
P. Filip
\vskip 5mm
{\small {\it Institute of Physics, Slovaka Academy of Sciences \\
D\'ubravsk\'a cesta 9, Bratislava 845 11, Slovakia}} \\
\end{center}
\vskip 5mm
\centerline{\bf Abstract}
The response of pseudoscalar and vector mesons to strong magnetic fields is studied 
within a simple constituent quark model using analogy with bound states of Positronium.
Magnetic moments of charged vector 
mesons $K^*$, $D^*$, $B^*$ are predicted and it is found
that $\eta$ mesons have magnetic polarizability.   
In extremely strong magnetic fields, behaviour of $J/\Psi$ mesons is discussed. 
We speculate on the existence of an induced
magnetic moment of $\eta$ meson.

\vskip 10mm
\section{\label{sec:intro}Introduction}
Since the pioneering suggestion of Gell-Mann and Zweig, 
that hadrons 
are composed of fractionally charged constituents, the quark model has
achieved a remarkable success. Chromo-electric charge (color) of $s$=1/2 quarks,
naturaly leads to the concept of chromo-magnetic moments, 
which are responsible for the "hyperfine" splitting of hadronic masses. 
Quarks, possessing the electric charge, should also
exhibit ordinary magnetic moment $\mu \approx \hbar Q/2m^*$
as inferred from Dirac equation.  
Indeed, experimentally measured magnetic moments of hyperons,
proton, neutron and $\Delta^{++}, \Delta^+$ resonaces \cite{PDG}
can be reasonably well understood 
as originating from quark magnetic moments \cite{ChaoPaper}.

One is tempted to ask then, whether quark-antiquark states (mesons) also
have magnetic moments. For charged $J^{P}$=$1^{-}$ mesons composed of different-flavour
$q\bar q$ pairs 
($\rho^{\pm},K^{\pm *}, D^{\pm *}$) the answer is simple and positive: Yes, such mesons should have a
magnetic moment. In the case of vector mesons with hidden flavour (J/$\Psi$, $\Upsilon$, $\phi$) a more detailed 
quantum approach is necessary. For pseudoscalar
$J^{P}$=0$^{-}$ mesons ($\pi, K, \eta$) the situation is even less clear. 
Can quantum
system with zero angular momentum have a magnetic moment? 

Based on the analogy with magnetic behavior of positronium ($e^+e^-$) 
and muonium ($\mu^+e^-$) triplet and singlet ground states,
we investigate here magnetic moments and polarizability of vector mesons J/$\Psi$, $\Upsilon$, $B^{*-}$
and magnetic polarizability of pseudoscalar mesons $\eta_c,\eta_b$ and $\eta$(547). Quenching of J/$\Psi$
decay in very strong magnetic fields (created in heavy ion collisions)
naturaly appears if our analogy with ortho-positronium is justified.

\section{Magnetic moments of quarks and baryons}

Assuming that each constituent quark
has magnetic moment $\mu_q = \frac{\hbar Q}{2m^*}$ 
(where $m^*$ is the effective quark mass), 
magnetic moments of  baryons can be calculated \cite{PDG} as: $\mu^* = \sum \mu_q$ for s=3/2 baryons
(e.g. $\Omega^-$ and $\Delta^{++}$, $\Delta^+$);
$\mu^* = \mu_{(s,c,b)}$ for baryons of type $\Lambda, \Lambda_c, \Lambda_b$ containing ($ud$) diquark
and different quark ($s,c,b$);
and $\mu^*$=$(4\mu_a$-$\mu_b)/3$ for type $(q_aq_b)$$q_a$ baryons (e.g. $n,p,\Xi$). 

Magnetic moments $\mu_q$ of quarks $u,d,s$ and their effective masses
inferred from the measured magnetic moments of hyperons, $p$ and $n$
are shown in the upper three rows of Table \ref{TabMagQuark}.

\begin{table}[ht]
\begin{center}
 \caption{Magnetic moments and masses of quarks.}
\begin{tabular}{|c|c|c|c|c|}
\hline
quark& Q& $\mu_q$ [$\mu_N$]& $m^*$\,\,[MeV]& $m_M$\, [MeV] \\
\hline
 u&\,2/3& 1.852& 338& 350\\
\hline
 d& -1/3& -0.972& 322& 370\\
\hline
 s& -1/3& -0.613& 510& 500\\
\hline
\hline
 c& \,2/3& 0.404& 1550& 1600\\
\hline
 b& -1/3& -0.066& 4730& 4770\\
\hline
 t& \,2/3& 0.004& 172900& - \\
\hline
\end{tabular}
\label{TabMagQuark}
\end{center}
\end{table}

Observing that effective ($m^*$) and constituent masses $m_M$ 
of $s,d,u$  quarks are similar,
one may predict magnetic moments of heavy  $c, b, t,$ quarks for which  
the corresponding heavy hyperon magnetic moments are not measured.

\section{Magnetic properties of mesons}
Bound states of quark-antiquark pairs (mesons) in the ground $S$-state
can have parallel or antiparallel spins of their constituents resulting in vector
($J^P$=1$^-$) or pseudoscalar ($J^{P}$=0$^{-}$) mesons. We shall assume here, that response
of these mesons to strong magnetic fields is similar to behavior
of positronium ($e^+e^-$) and muonium  ($e^-\mu^+$) in the magnetic field.
For example, triplet S-state of positronium (with paralel $e^+e^-$ spins) is analogous
to the quantum state of vector mesons $\varphi$(1020), $\Upsilon$ or $J/\Psi$ with $J^{PC} = 1^{--}$,
while the singlet state of positronium (antiparallel $e^+e^-$ spins) resembles
the structure of pseudoscalar $\eta_b,\eta_c$ mesons. 

For vector mesons composed of unlike-flavour quark-antiquark pair with parallel spins
e.g. $\rho^+$($u\bar d$), $K^{*-}(s\bar u)$ or $D^{*+}(c\bar d)$,
one can use analogy with muonium ($e^-\mu^+$) 
and add 
the magnetic moment of quark and antiquark $\mu_{q\bar q}$=$|\mu_q|$+$|\mu_{\bar q}|$.
This approach gives $\mu$=-2.82$\mu_N$ and 2.46$\mu_N$, -1.37$\mu_N$, -1.02$\mu_N$, -1.92$\mu_N$
for mesons $\rho^-(d\bar u)$ and $K^{*+}(u\bar s)$, $D^{*-}(d\bar c)$, $D^{*-}_s(s\bar c)$,  $B^{*-}(b\bar u)$.
Our obtained value $\mu_{\rho^-}$=-2.82$\mu_N$ agrees well with lattice calculations \cite{magMomRho}.

\begin{figure} [h,t]           
\begin{minipage}{15pc}
\includegraphics[width=15pc]{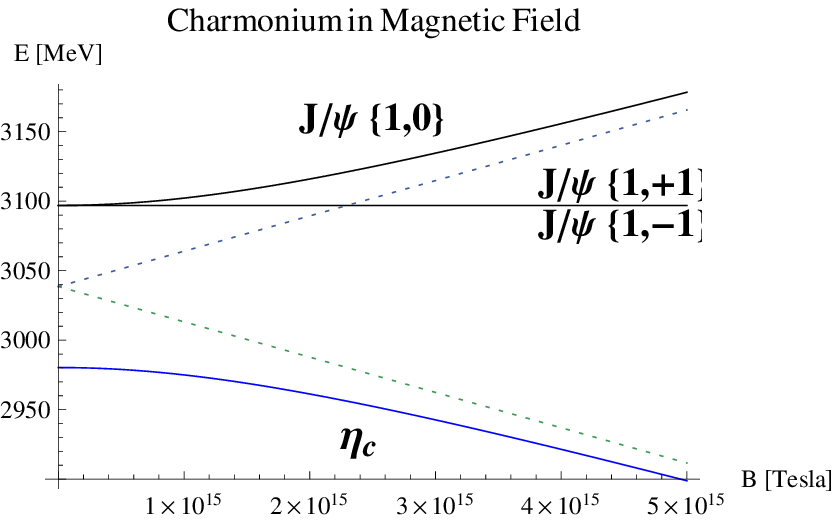}   
\vspace{-2mm} \caption{\label{Fig1}Energy of $\eta_c$ and $J/\Psi$ in very strong magnetic fields.}
\end{minipage}\hspace{2pc}
\begin{minipage}{15pc}
\includegraphics[width=15pc]{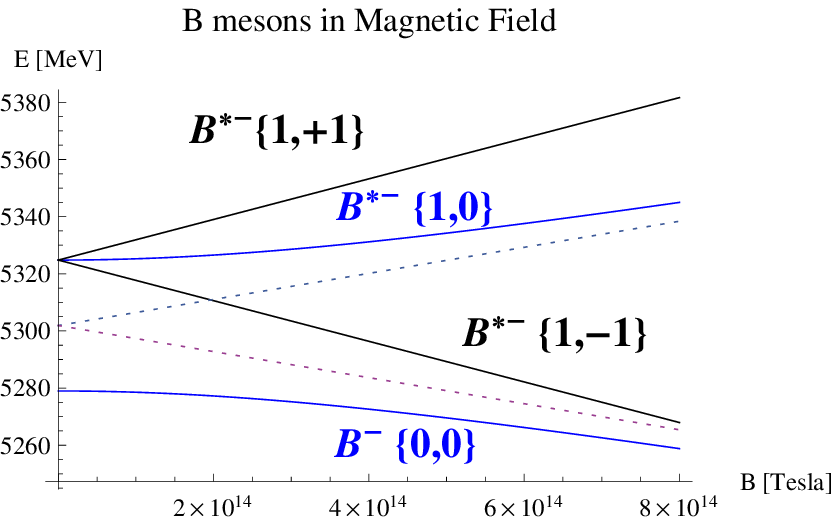}
\caption{\label{Fig2}Energy of $B^-(b\bar u)$ and $B^{*-}$ in very strong magnetic fields. }
\end{minipage} 
\label{FigX1}
\end{figure}

Using the analogy with triplet and singlet states of Positronium  in magnetic field \cite{Positronium}, one can predict 
that $J/\Psi$ and $\eta_c$ mesons 
do not have magnetic moment (see Fig.\ref{Fig1}), which applies also to 
$\Upsilon(b\bar b)$, $\phi(s\bar s)$, $\eta_b$,  and $\eta$, $\eta'$ mesons. 
A possibility of the magnetic quenching of $J/\Psi$ decay (as observed for ortho-positronium \cite{QuenchEe}) is very interesting
and it deserves a detailed study.
Energy of the singlet muonium state in magnetic field behaves
as \cite{MuoniumBook}
\begin{equation}
E_{(\mu^+ e^-)} = -
 \frac{\Delta E_{h\!f}}{2}\sqrt{1+\Big[\frac{(|\mu_e|+|\mu_\mu|)B}{\Delta E_{h\!f}}\Big]^2}
\label{Qe6}
\end{equation}
which decreases as $E^-\approx -\Delta E_{h\!f}/2 - (\tilde\mu' B)^2/\Delta E_{h\!f}$
for small $B$ fields (here $\tilde\mu'$=($|\mu_e|$+$|\mu_\mu|$)/2).
Muonium singlet state thus achieves induced magnetic moment 
$\tilde\mu[B]=\tilde \mu'\cdot\kappa(B)$, where $\kappa(B)=\tilde\mu'B/\Delta E_{h\!f}$ and
$\Delta E_{h\!f}$=1.8$\cdot$10$^{-5}$ eV.

Replacing magnetic moments and masses
of $e^-$ and $\mu^+$ in Eq.(\ref{Qe6}) 
by corresponding quark values (from Tab. 1) and using $\Delta E_{h\!f}$=45.8 MeV, one can predict also the magnetic behavior
of  $B^{*-}$ and $B^-(b\bar u)$ mesons (see Fig.2).

Two ($m_z$=$\pm 1$) components of $B^{*-}$ triplet state do have magnetic moment 
$\tilde\mu$=$\pm$($|\mu_b|$+$|\mu_{\bar u}|$), while $m_z$=$\,0$ component of $B^{*-}$  and the meson $B^-(b\bar u)$  have
magnetic polarizability $\beta$. 
Using $\Delta E$=$\,-\beta B^2 (2\pi/\mu_o)$  from \cite{Schumacher} and comparing to Eq.(\ref{Qe6})
 one has $\beta_{b\bar u} = \frac{\mu_o}{2\pi}\cdot \frac{(|\mu_b|-|\mu_{\bar u}|)^2/4}{\Delta E_{h\!f}}$, which
gives $\beta$=$5.5\cdot 10^{-4}$fm$^3$ for $B^{-}$.

Quadratic energy response of $\eta_c (2981)$ to magnetic field (Fig.\ref{Fig1}) suggests magnetic polarizability
$\beta = 0.44\cdot 10^{-4}$fm$^3$, obtained using $\Delta E_{h\!f}$=116.6MeV.  
For $\eta_b (9391)$ one can predict $\beta = 0.02\cdot 10^{-4}$fm$^3$ (here $\Delta E_{h\!f}$= 69.3 MeV).

\section{Induced magnetic moment of $\eta$ meson}
We may now suggest, that $\eta$(547) also has magnetic polarizability, 
due to the similarity of its quantum structure with $\eta_c$. Assuming $\omega(782)$ meson to
be spin-triplet partner of $\eta$(547) one has  
$\Delta E_{h\!f}$=235$\,$MeV and this gives magnetic polarizability 
1.3$\cdot$10$^{-4}$ $< \beta <$ 4.6$\cdot$10$^{-4}\,$fm$^{-3}$ for $\eta$(547),
depending on the exact nature of its ($c_1 u\bar u$+$c_2 d\bar d$) quantum state. 
Analogously to Positronium 
(see text below Eq.(\ref{Qe6})), meson $\eta $(547) in magnetic fields
should behave as having an induced magnetic moment  
$\tilde\mu[B]$=$\tilde \mu'$$\cdot$$\kappa(B)$, where $\kappa(B)$=$\tilde\mu'B/\Delta E_{h\!f}$ and
$\tilde\mu'$=($|\mu_{u(d)}|$+$|\mu_{\bar u(\bar d)}|$)/2.
However, it remains a question,
whether real dipole magnetic field is generated by such induced magnetic moment. 

From nuclear physics we know, that
in H$^3$ nucleus two neutrons form S=0 scalar state (similar to singlet-positronium) and the
magnetic moment of H$^3$ is to be generated by the proton with $\mu_p = 2.79 \mu_N$. 
However,
$\mu_{H^3} = 2.98\mu_N$, which is 7\% larger compared to $\mu_p$. Where does 7\% increase come from?

Since two neutrons in $S$=0 state are located in the magnetic field of the proton in H$^3$,  
one can speculate, that the bound state of two neutrons does have magnetic polarizability $\beta$, and 
that induced magnetic moment $\mu^*$ of scalar di-neutron contributes by 7\% to the magnetic moment
of H$^3$ nucleus.  

If such picture is correct, then induced magnetic moment $\mu^*$ of $\eta$(547) meson can produce its own magnetic field - it is real.

Consequently, also scalar diquarks in baryons, e.g.$\,(ud)$ diquark in $\Lambda$, $\Lambda_c$  hyperons or Nucleon, may contribute
to the total observed magnetic moment.

\vskip 1.5mm
{\bf Acknowledgements:} This work has been supported by Grant agency VEGA(1/0171/11) 
and APVV-0177-11 in Slovakia and by JINR Dubna.

\end{document}